\documentclass[12pt]{article}
\usepackage{amssymb}
\def\beq{\begin{equation}}
\def\eeq{\end{equation}}
\def\ds{\displaystyle}
\def\ov{\overline}

\def\wh{\widehat}
\usepackage[all,2cell,dvips]{xy}
\def\IR{\relax{\rm I\kern -.18em R}}

\begin{document}
\title{ The Green's Function of the Supersymmetric $D=1$ Heat Equation }
\author{ \Large S. Andrea*, A. Restuccia**, A. Sotomayor***}
\maketitle{\centerline {*Departamento de Matem\'{a}ticas,}}
\maketitle{\centerline{**Departamento de F\'{\i}sica}}
\maketitle{\centerline{Universidad Sim\'on Bol\'{\i}var}}
\maketitle{\centerline{***Departamento de Ciencias B\'{a}sicas}}
\maketitle{\centerline{Unexpo, Luis Caballero Mej\'{\i}as }}
\maketitle{\centerline{e-mail: sandrea@usb.ve, arestu@usb.ve,
asotomay@zeus.unexpo.edu.ve }}

\begin{abstract}A rigorous treatment is given of the Green's function of the
$N=1$ supersymmetric heat equation in one spatial dimension with a
distribution initial value. The asymptotic expansion of the
supersymmetric Green's function as $t$ tends to $0^+$ is also
derived. The coefficients of the expansion generate all the
members of the supersymmetric $N=1$ KdV hierarchy.

\end{abstract}

\section{Introduction} We analize in this work the quantum
mechanical problem, in the ``euclidean" domain, associated with
supersymmetric integrable models. The $N=1,2$ supersymmetric
extensions of the KdV equation were found several years ago
\cite{M,Manin,Laberge,Popowicz}.

Later, the $N=3,4$ cases were also discussed  \cite{Yung,Krivonos,
Delduc}.

A lot of work has been done on the subject, in particular the
bihamiltonian structure of the super KdV equations was studied in
\cite{Oevel,Farrill}. For a review of the $N=1,2$ supersymmetric
KdV equations see \cite{Mathieus}. Extensions of the
supersymmetric model were proposed in \cite{Andrea}.
Supersymmetric quantum problems in other contexts are also
relevant. In particular, the quantum mechanical problem arising
from the regularization of the Supermembrane in eleven dimensions.
The nature of the spectrum of the corresponding hamiltonian has
been described in \cite{de Wit, B. de Wit, Boulton, Alvaro},
however a satisfactory understanding of the corresponding Green's
function has not been achieved.

In \cite{Adrian}, using the exact sequence for the $N=1$
supersymmetric ring, explicit expressions for the Green's function
of the $N=1$ supersymmetric heat equation were obtained and using
a formal asymptotic expansion of it, when $t\rightarrow0^+$, all
the supersymmetric $N=1$ KdV hierarchy was generated. In the
present work we prove the convergence of the series defining the
supersymmetric Green's function with a distribution initial value
and give a rigorous treatment of its asymptotic expansion as
$t\rightarrow0^+$.

We thus complement our previous work with rigorous statements. In
section 2 we present general considerations, the new results are
developed in sections 3 and 4.

\section{The Susy Heat Operator}
The usual homogeneous heat equation with potential may be
written
$$
Hu = \phi u
$$
with $H = \partial_t-\partial^2_x$ and $u(t,x)$, $\phi(x)$
real--valued functions of real variables.  In the supersymmetric
extension to be considered here, the functions will take their
values in an exterior algebra $\Lambda$.  In more detail:

Let $\Lambda$ be generated by anticommuting elements
$e_1,\ldots,e_m$ which satisfy $e^2_i = 0$. The subspaces of
bosonic and fermionic elements of $\Lambda$ are the $\pm 1$
eigenspaces of the involution $\lambda\rightarrow\bar{\lambda}$
wich sends the generators $e_i$ to $-e_i$.

For $1\leq i\leq m$ the operator
$\partial_i:\Lambda\rightarrow\Lambda$ is defined by the equations
$\partial_i\lambda=e_i$ and $\lambda=e_i\alpha+\beta$, the
elements $\alpha$ and $\beta$ being independent of the generator
$e_i$. This operator is a superderivation because it anticommutes
with the involution and satisfies the product rule
$\partial_i(\lambda\mu)=(\partial_i\lambda)\mu+\bar{\lambda}(\partial_i\mu).$

Let $\Lambda_1\subset\Lambda$ be the subalgebra generated by
$e_1,\ldots e_n$ where $n<m$, and let
$\Theta=\left\{e_{n+1},\ldots,e_m\right\}$ be the set of remaining
generators. If $\theta\in\Theta$ then the superderivation
$\partial_\theta:\Lambda\rightarrow\Lambda$ also acts in the ring
of all functions $f:\mathbb{R}\rightarrow\Lambda$.

If the elements of $\Theta$ are written as
$\theta_1,\theta_2,\ldots$ then the functions
$f:\mathbb{R}^p\rightarrow\Lambda$ can be written in the notation
\[f(x,\theta)=\sum\theta_{j_1}\cdots\theta_{j_q}f_J(x)\] with
$J=(j_1<\cdots <j_q)$ and $f_J:\mathbb{R}^p\rightarrow\Lambda_1$.
One sees that $\partial_\theta f$ has the same $f_J$ but with
different coefficients. Evidently $\Theta$ could have any size;
but in what follows it will have no more than three elements
$\theta,\theta^\prime,\theta^{\prime\prime}$.

Supposing now that $p=1,n+1=m$ and $\Theta=\left\{\theta\right\}$
we begin by observing that the ring of functions
$f:\mathbb{R}\rightarrow\Lambda$ admits the operator

$$
D = \partial_\theta + \theta\partial_x,
$$
at least if we restrict to the case $f(x,\theta) = f_1(x) + \theta
f_2(x)$ with differentiable $f_1$ and $f_2$.  Then $Df = f_2(x) +
\theta\partial_xf_1(x)$.

The functions $u(t,x)$ and $\phi(x)$ in the usual heat equation
are now replaced by
$$
u(t,x,\theta) = u_1(t,x) + \theta u_2(t,x)
$$
$$
\phi(x,\theta) = \alpha(x) + \theta\beta(x)
$$
with $u_1$, $u_2$, $\alpha$, and $\beta$ taking their values in
$\Lambda_1$.  The SUSY heat equation, with the same $H =
\partial_t-\partial^2_x$ as before, is defined to be
$$
Hu = D(\phi u).
$$
In terms of the components it is
$$
Hu_1 = \beta u_1 + \ov\alpha u_2
$$
$$
Hu_2 = \partial_x(\alpha u_1).
$$
The usual heat equation $Hu_1 = \beta u_1$ reappears if $\alpha
\equiv 0$ and the values of $\beta$ and $u_1$ are multiples of the
identity element of $\Lambda$.  This requires $\beta$ to be of
even type, that is, $\ov\beta = \beta$.  This possibility is
admitted by assuming that the SUSY potential function is
fermionic, that is, $\ov\phi = -\phi$.  Then $\ov\alpha =
-\alpha$, and the componentwise version of the SUSY heat equation
is
$$
Hu_1 = \beta u_1 - \alpha u_2
$$
$$
Hu_2 = \partial_x(\partial u_1).
$$

Going back for a moment to the usual heat equation we recall that
the inhomogenous equation
$$
Hu = \phi u+f,
$$
with $f(t,x)$ defined in the upper half plane $t > 0$, is solved
by
$$
u(t,x) = \int^t_0 \int^\infty_{-\infty} g(t-s,x,y)f(s,y)dyds
$$
where $g(t,x,y)$ is the Green's function of the equation:  it
satisfies $Hg =\phi g$ as a function of $x$ and $t$ when $t > 0$,
and has the initial value
$$
\lim_{t\downarrow 0} g(t,x,y) = \delta(x-y),
$$
the distributional Dirac delta function.

An analogous SUSY Green's function $G$ is now to be developed for
solving the SUSY equation $H u = D(\phi u)+Df$. There are three
steps:  first a SUSY integration formula is developed for solving
$Hu = Df$; second an explicit construction is given for the SUSY
Green's function $G$ with attention to its symmetries; and third
it is shown that $G$ has an asymptotic expansion in powers of $t$.

\medskip
Guided by the usual $g(t,x,y)$ we might suggest that the SUSY
Green's function should depend $(x,\theta)$ and $(y,\theta')$ in
place of $x$ and $y$.  Accordingly
$$
G(t,x,y;\theta,\theta') = a + \theta b + \theta'c + \theta\theta'd
$$
where $a$, $b$, $c$, and $d$ are $\Lambda_1$--valued functions of
$t$, $x$ and $y$,  Then, given $f(t,x,\theta)$, an analogous
definition of $u(t,x,\theta)$ would be
$$
u(t,x,\theta) = \int^t_0 \int^\infty_{-\infty}
       G(t-s,x,y;\theta,\theta')f(s,y,\theta')d\theta'dyds.
$$
The integration with respect to $\theta'$ is to be understood as
the partial derivative with respect to $\theta'$: that is, given
in general any $J(x_1,\ldots,x_p;\theta_1,\ldots,\theta_n)$, its
integral with respect to $\theta_j$ is defined to be
$$
\int J d\theta_j = \frac{\partial}{\partial\theta_j} J.
$$
With $f(t,x,\theta) = f_1(t,x) + \theta f_2(t,x)$, the integration
$d\theta'$ leaves the integrand in the form
$$
(cf_1 + \ov af_2) + \theta(-df_1 - \ov bf_2).
$$
Therefore, with $u(t,x,\theta) = u_1(t,x) + \theta u_2(t,x)$ we
have
$$
\begin{array}{l}
u_1 = \int cf_1 + \int \ov af_2 \\
u_2 = -\int df_1 - \int \ov bf_2
\end{array}
$$
with $\int$ signifying the remaining integrations $dyds$.

Since $Df = f_2+\theta\partial_xf_1$, the equation $Hu = Df$ will
be satisfied if we choose $b = c = 0$, while $a$ and $d$ should
satisfy
$$
\begin{array}{l}
H(\textstyle\int \ov a f_2) = f_2 \\
H(\textstyle\int df_1) = -\partial_xf_1.
\end{array}
$$
Therefore $a$ should be the standard Gauss bell heat kernel and
$d$ should be the negative of its derivative with respect to $x$.
With the definition
$$
G_0(t,x-y;\theta,\theta') = \frac{1}{\sqrt{4\pi t}}
      e^{-\frac{(x-y)^2}{4t}} \Big(1 + \theta\theta'\frac{x-y}{2t}\Big)
$$
we have a SUSY Green's function which solves $Hu = Df$ by $u =
\int G_0f$.

The defining properties of $G_0$ are that it should satisfy $HG_0
= 0$ for $t > 0$, with initial value
$$
\lim_{t\downarrow 0} G_0 = \delta(x-y) -
\theta\theta'\delta'(x-y),
$$
with $\delta'$ being the distributional derivative of the Dirac
delta function.

In addition one observes that $G_0$ is bosonic $(\ov G_0 = G_0)$
and enjoys the symmetry
$$
G_0(t,x,y;\theta,\theta')
    = G_0(t,y,x;\theta',\theta).
$$

Turning now to the equation
$$
Hu = D(\phi u)+Df
$$
with a fermionic potential $\phi(x,\theta) =
\alpha(x)+\theta\beta(x)$, we ask for its Green's function.
Following what was seen for $G_0$, we should ask that
$G(t,x,y;\theta,\theta')$ satisfy $HG = D(\phi G)$ as a function
of $t$, $x$, and $\theta$ when $t > 0$, and that it should have
initial value $\ds\lim_{t\downarrow 0}G =
\delta(x-y)-\theta\theta'\delta'(x-y)$. Supposing that such a $G$
exists, we argue that $u = \int Gf$ satisfies $Hu = D(\phi u)+Df$.

Writing
$$
G = A+\theta B + \theta'C + \theta\theta'D
$$
with $A$, $B$, $C$, and $D$ functions of $t$, $x$, and $y$, we
note that the componentwise version of $HG = D(\phi G)$ is
$$
\begin{array}{l}
HA = \beta A - \alpha B \\
HB = \partial_x(\alpha A) \\
HC = \alpha D + \beta C \\
HD = -\partial_x(\alpha C).
\end{array}
$$
(Rewritten with explicit independent variables the first equation,
for example, is
$$
(\partial_t-\partial^2_x)A(t,x,y)
     = \beta(x)A(t,x,y) - \alpha(x)B(t,x,y).)
$$
The parities $\ov A = A$, $\ov D = D$, $\ov B = -B$ and $\ov C =
-C$ give
$$
\begin{array}{l}
u_1 = \int Cf_1 + \int Af_2 \\
u_2 = \int Bf_2 - \int Df_1.
\end{array}
$$
To apply $H$ we use the general fact that
$$
v(t,x) = \int^t_0 \int^\infty_{-\infty} K(t-s,x,y)\ell(s,y)dyds
$$
implies
$$
Hv(t,x) = \int^t_0 \int^\infty_{-\infty}
(HK)(t-s,x,y)\ell(s,y)dyds
     + \lim_{s\downarrow 0} \int^\infty_{-\infty} K(s,x,y)\ell(t,y)dy.
$$
The initial values imposed on $G$ leave
$$
\begin{array}{l}
Hu_1 = f_2 + \int (HC)f_1 + \int (HA)f_2 \\
Hu_2 = \partial_xf_1 + \int (HB)f_2 - \int (HD)f_1.
 \end{array}
$$
However
$$
\begin{array}{c} \int (HA)f_2
    = \int^t_0 \int^\infty_{-\infty} \big(B(x)A(t-s,x,y)f_2(s,y)
            - \alpha(x)B(t-s,x,y)f_2(s,y)\big)dyds \\
   = \beta \int Af_2 - \alpha \int Bf_2.
\end{array}
$$
In like manner
$$
\begin{array}{c} \int (HB)f_2
    = \int^t_0 \int^\infty_{-\infty} \partial_x\alpha(x)A(t-s,x,y)
        f_2(s,y)dyds \\
   = \partial_x\alpha \int Af_2.
\end{array}
$$
Similar computations for $HC$ and $HD$ lead to
$$
\begin{array}{l}
Hu_1 = f_2 + \beta u_1 - \alpha u_2 \\
Hu_2 = \partial_x f_1 + \partial_x(\alpha u_1),
\end{array}
$$
which is just the componentwise version of $Hu = D(\phi u)+Df$, as
desired.

\bigskip
\section{Construction of the SUSY Green's Function}

\medskip
A potential $\phi(x,\theta) = \alpha(x) + \theta\beta(x)$ is
given.  In addition to the conditions $\ov\alpha = -\alpha$ and
$\ov\beta = \beta$, we assume that $\alpha$ and $\beta$ are
bounded infinitely differentiable functions from $\Bbb R$ to
$\Lambda$, and that all their derivatives are bounded.

Formally we put
$$
G = \sum^\infty_{m=0} G_m
$$
where $G_0$, defined in the preceding section, satisfies $HG_0 =
0$ in the upper half plane and has the initial value
$\delta(x-y)-\theta\theta'\delta'(x-y)$.

The desired equation $HG = D(\phi G)$ suggests the recursion
$H_{m+1} = D(\phi G_m)$.  This equation in turn is solved by
integration with $G_0$, as follows.

For fixed $y$, $\theta'$ the function $\phi G_m$ takes the form
$$
f(t,x,\theta) = \phi(x,\theta)G_m(t,x,y;\theta,\theta')
$$
for variable $x$, $t$, $\theta$.  To solve $Hu = Df$ we replace
$x$, $t$ and $\theta$ by variables of integration $u$, $s$ and
$\theta''$.  Then
$$
u(t,x,\theta)
    = \int^t_0 \int^\infty_{-\infty} G_0(t-s,x-u;\theta,\theta'')
        f(s,u,\theta'')d\theta'' duds,
$$
giving
$$
G_{m+1}(t,x,u;\theta,\theta')
      = \int^t_0\!\int^\infty_{-\infty} G_0(t-s,x-u;\theta,\theta'')
         \phi(u,\theta'') G_m(s,u,y;\theta'',\theta')d\theta'' duds.
$$

We may write
$$
G_m = A_m + \theta B_m + \theta'C_m + \theta\theta'D_m
$$
in general, where for $m = 0$ we already know that $G_0 =
g-\theta\theta'h$, with $g(t,x-y) = \frac{1}{\sqrt{4\pi t}}
e^{-\frac{(x-y)^2}{4t}}$ and $h(t,x-y) = \frac{\partial}{\partial
x} g(t,x-y)$.

The formal integration $d\theta''$ is done by applying
$\frac{\partial}{\partial\theta''}$ to
$$
(g-\theta\theta''h)(\alpha + \theta''\beta)(A_m + \theta''B_m
        + \theta'C_m + \theta''\theta'D_m).
$$
This leaves
$$
(g\beta A_m-g\alpha B_m) + \theta(h\alpha A_m)
       + \theta'(g\beta C_m + g\alpha D_m) - \theta\theta'(h\alpha C_m).
$$
Therefore except for the remaining integrations $du ds$ we have
the componentwise recursion formula:
$$
\begin{array}{l}
A_{m+1} = g\beta A_m - g \alpha B_m \\
B_{m+1} = h\alpha A_m \\
C_{m+1} = g\beta C_m + g\alpha D_m \\
D_{m+1} = -h\alpha C_m.
\end{array}
$$
This can be recast in matrix form as

\[\left(\begin{array}{cc}  A_{m+1}  &-C_{m+1} \\ B_{m+1}  &D_{m+1}
\end{array}\right)
   = \left(\begin{array}{cc}  g\beta  &-g\alpha \\ h\alpha  &0
   \end{array}\right)
    \left( \begin{array}{cc} A_m  &-C_m  \\ B_m  &D_m \end{array}\right),\]

giving at least formally the explicit formula
$$
\left(\begin{array}{cc}  A_m  &-C_m \\  B_m  &D_m
\end{array}\right)
    = {\left(\begin{array}{cc}  g\beta  &-g\alpha \\ h\alpha  &0 \end{array}\right)}^m
      \left(\begin{array}{cc}  g  &0 \\  0  &-h \end{array}\right) .
$$
The remaining integrations of the real variables $u$ and $s$ must
now be examined.  The equation $B_1 = h\alpha g$ for example means
that
$$
B_1(t,x,y) = \int^t_0 \int^\infty_{-\infty} h(t-s,x-u)\alpha(u)
        g(s,u-y)duds.
$$

The general case involves various products of the four functions
$\alpha$, $\beta$, $g$, and $h$.  To treat it one can define a
``word'' to be any element $W$ of a finite Cartesian product $E
\times \Omega \times E \times \Omega \times \cdots \times \Omega
\times E$ where $E = \{g,h\}$ and $\Omega = \{\alpha,\beta\}$.
Then, given
$$
W_m = e_0 \omega_1 e_1\omega_2 \cdots \omega_me_m,
$$
a function of $t$, $x$ and $y$ arises through the integral
$$
W_m(t,x,y) = \int_{I_m(t)} e_0(t-s_1,x-u_1)\omega_1(u_1)
       e_1(s_1-s_2u_1-u_2)\cdots \omega_m(u_m)e_m(s_m,u_m-y)
$$
where $I_m(t) \subset \Bbb R^{2m}$ is the set of
$(s_1,u_1,s_2,u_2,\ldots,s_m,u_m)$ satisfying $t > s_1 > s_2 >
\cdots > s_m > 0$.  The convergence of this integral and of the
series defining $G$ are now to be discussed, as well as the
distribution initial value of $G$ as $t$ tends to zero.

First we observe that $g$ and $h$ are in the space $ \mathcal{K}$
of real--valued continuous functions $k(x,t)$ in the upper half
plane which satisfy, for all $0 < T_1 <
T_2$,\[\begin{array}{c}\int_0^{T_1}\int_{-\infty}^{\infty}|k(s,u)|duds<\infty
\\ \begin{array}{l}\sup\hspace{3mm}|k(s,u)|<\infty. \\\small\begin{array}{l}-\infty<u<\infty
\\ T_1<s<T_2 \end{array}
 \end{array}
\end{array}\]

The space $\mathcal{K} $ is a commutative ring under convolution
$$
(k_1*k_2)(t,x) = \int^t_0 \int^\infty_{-\infty} k_1(t-s,x-u)
         k_2(s,u)duds.
$$
If we take $k(t,x)$ to be some positive multiple of $g(t,x) +
|h(t,x)|$, we obtain the estimate
$$
|W_m(t,x,y)| \leq l(t,x-y)
$$
in which $l$ is the $(m+1)$--fold convolution
$$
l = k*\cdots *k.
$$
This shows that $W_m(t,x,y)$ is in $ \mathcal{K}$ for any value of
$y$, and is defined by an absolutely convergent integral.

In order to estimate $l$ in points of the strip $0 < t < 1$ we
define the three auxiliary functions

\[k_1(t) = \int^t_0 \int^\infty_{-\infty} k(s,u)duds\]
\[\small\begin{array}{r}\tau(t)=\sup\hspace{3mm}k(s,u)\\-\infty<u<\infty
\\t<s<1 \end{array}\]
\[\small\begin{array}{r}\rho(r)=\sup\hspace{3mm}k(s,u)\\u<-r\mathrm{\:or\:} u>r
\\0<s<1 \end{array}\]

Then from the definition of the $(m+1)$--fold convolution one can
deduce
$$
\begin{array}{c}
l(x,t) \le (m+1)(k_1(t))^m \tau\Big(\frac{t}{m+1}\Big) \\
l(x,t) \le (m+1)(k_1(t))^m \rho\Big(\frac{|x|}{m+1}\Big).
\end{array}
$$
For some $C$ the function $k(t,x)$ satisfies
$$
k(t,x) \le \frac{C}{x^2+t}
$$
in the strip $0 < t < 1$, as can be seen by using $e^r >
1+r+\frac{r^2}{2}$ in the formulas for $g$ and $h$.  Hence
$\tau(t) \le \frac Ct$ and $\rho(r) \le \frac{C}{r^2}$. Also
$k_1(t) \le C\sqrt t$, again by checking the same formulas.

This gives us the estimates
$$
\begin{array}{c}
l(x,t) \leq (m+1)^2 C^{m+1} t^{\frac{m-2}{2}} \\
l(x,t) \leq (m+1)^3 C^{m+1} \frac{t^{\frac m2}}{x^2}.
\end{array}
$$
The first estimate gives the absolute convergence of any series
$\ds\sum^\infty_1 W_m(t,x,y)$ in some strip, say $0 < t <
\frac{1}{2C^2}$.  Taking off the first two terms we also see that
$\ds\lim_{t\rightarrow 0} \sum^\infty_{m=3} W_m = 0$, uniformly in
$x$ and $y$.

In order to show that the full series satisfies
$\ds\lim_{t\rightarrow 0} \sum^\infty_{m=1}W_m = 0$ in the
distribution sense we must examine $W_1$ and $W_2$.

The second estimate shows, for any $m \ge 1$, that
$\ds\lim_{t\rightarrow 0} W_m(t,x,y) = 0$ uniformly in any set of
$x$ and $y$ of the form $|x-y| \ge r > 0$.  Therefore the desired
initial value of $\ds\sum^\infty_{m=1} W_m$ would follow if $W_1$
and $W_2$ were bounded functions.

But this is clear for $W_2$, by the first estimate. For $W_1$ we
note that the recursion formula for $G_1$ only mentions $g\beta
g$, $g\alpha h$, and $h\alpha g$; therefore $|h|*|h|$ need not be
analyzed.

One can easily check
$$
\begin{array}{c}\Big(\int^\infty_{-\infty} |g(s,u)|^2du\Big)^{1/2}
     = C_1 s^{-\frac 14} \\
\Big(\int^\infty_{-\infty} (h(s,u))^2du\Big)^{1/2}
    = C_2s^{-\frac 34}
\end{array}
$$
for some constants, giving
$$
\int^\infty_{-\infty} g(t-s,x-u)|h(s,u|du
     \le C_1C_2(t-s)^{-\frac 14} s^{-\frac 34}
$$
by the Schwartz inequality.  Then because $\int^t_0 (t-s)^{-\frac
14} s^{-\frac 34}ds$ is a finite positive constant independent of
$t > 0$, we conclude that $g*|h|$ is a bounded function for
$-\infty < x < \infty$, $t > 0$.  (The remaining case $g*g =
tg(x,t)$ is obvious.)

To sum up: $\ds\sum^\infty_1 W_m(t,x,y)$ is absolutely and
uniformly convergent in some strip $0 < t < \varepsilon$, and has
distribution limit zero as $t$ tends to zero.

 From the recursion formulas for the $G_m$ it is clear that each
of $A_m,\ldots,D_m$ is a linear combination of no more than $2^m$
words $W_m$, with $\pm 1$ coefficients.  Therefore the SUSY
Green's function is given by the series
$$
G(t,x,y;\theta,\theta') = \sum^\infty_0 G_m,
$$
convergent in a smaller strip $0 < t < \varepsilon/2$, with the
same initial value as the first term $G_0$:
$$
\lim_{t\rightarrow 0} G = \delta(x-y) - \theta\theta'\delta'(x-y)
$$

The construction of $G$ being complete, we turn now to its
symmetries. These depend on the words produced by the recursion
algorithm.  If $W_m = e_0\omega_1e_1\omega_2 \cdots \omega_me_m$.
Then
$$
W_m(t,x,y) = \pm W^*_m(t,y,x)
$$
where
$$
W^*_m = e_m\omega_m \cdots \omega_2e_1\omega_1e_0
$$
is the word $W_m$ written backwards, and the sign depends on how
many times $h$ and $\alpha$ appear in the word.  Here is the
proof:

For fixed $t > 0$ the involution of $\Bbb R$ given by $s
\rightarrow \ov s = t-s$ produces a volume--preserving self--map
of the set $I_m(t) \subset \Bbb R^{2m}$ where the integration
takes place: $(s_1,u_1,s_2,u_2,\ldots,s_mu_m)$ is sent to $(\ov
s_m,u_m,\ldots,\ov s_1,u_1)$. After this change of variable the
new integrand is the product of $e_0(t-\ov s_m,x-u_m)e_1(\ov
s_m-\ov s_{m-1},u_m-u_{m-1})\cdots e_m(\ov s_1,u_1-y)$ and
$\omega_1(u_m)\cdots \omega_m(u_1)$.

At this point we note that $g$ and $h$ satisfy $g(s_1-s_2;u_1-u_2)
= g(\ov s_2-\ov s_1,u_2-u_1)$ and $h(s_1-s_2,u_1-u_2) = -h(\ov
s_2-\ov s_1,u_2-u_1)$ when $t > s_1 > s_2 > 0$.  Consequently the
products of $g$'s and $h$'s before and after the change of
variable differ by the sign $(-1)^\mu$, $\mu$ being the number of
times $h$ appears among $\{e_0,e_1,\ldots,e_m\}$.

On the other hand the products of $\alpha$'s and $\beta$'s satisfy
$$
\omega_1(u_n) \cdots \omega_n(u_1)
     = (-1)^{\frac{\nu(\nu-1)}{2}} \omega_n(u_1) \cdots \omega_1(u_n),
$$
$\nu$ being the number of times the fermionic term $\alpha$
appears among $\{\omega_1,\ldots,\omega_m\}$.  The factors $e_k$
commute with all elements of the exterior algebra, so after one
more reordering we obtain
$$
W_m(t,x,y) = W^*_m(t,y,x)(-1)^\mu (-1)^{\frac{\nu(\nu-1)}{2}}.
$$
The quantity $\varepsilon(W) = (-1)^\mu
(-1)^{\frac{\nu(\nu-1)}{2}}$ is well--defined for any word in the
four--letter alphabet $\{g,h,\alpha,\beta\}$ and satisfies
$$
\varepsilon(W_1W_2) =
\varepsilon(W_1)\varepsilon(W_2)(-1)^{\nu_1\nu_2}.
$$
The recursion relation
$$
\left(\begin{array}{cc} A_{m+1}  &-C_{m+1} \\ B_{m+1}  &D_{m+1}
\end{array}\right)
  = \left(\begin{array}{cc} g\beta A_m-g\alpha B_m &-g\beta C_m-g\alpha D_m \\
         h\alpha A_m   &-h\alpha C_m \end{array}\right)
$$
gives
$$
\left(\begin{array}{cc}  A_1  &-C_1 \\ B_1  &D_1
\end{array}\right) =
\left(\begin{array}{cc} g\beta g &g\alpha h \\ h\alpha g  &0
\end{array}\right)
$$
and
$$
\left(\begin{array}{cc} A_2  &-C_2 \\ B_2  &D_2 \end{array}\right)
   = \left(\begin{array}{cc}  g\beta g\beta g - g\alpha h\alpha g  &g\beta g\alpha h\\
            h\alpha g\beta g    &h\alpha g\alpha h   \end{array}\right).
$$

Inspection shows that $\varepsilon(A_m) = +1$ while
$\varepsilon(B_m) = \varepsilon(C_m) = \varepsilon(D_m) = -1$ for
$m = 1$ and 2, and that $A_m$ and $D_m$ are bosonic while $B_m$
and $C_m$ are fermionic, this referring to the parity of the
number of times $\nu$ that the letter $\alpha$ appears in the
word.

The recursion relation can be used to show that these statements
hold for all $m$, namely, that $B_m$ and $C_m$ are formal linear
combinations of fermionic words satisfying $\varepsilon = -1$,
while $A_m$ and $D_m$ are combinations of bosonic words with
$\varepsilon = 1$, resp. $\varepsilon = -1$, for the terms in the
two cases.  (For example $\varepsilon(h\alpha) = -1$,
$\varepsilon(g\alpha h) = -1$ giving $\varepsilon(h\alpha g\alpha
h) = -\varepsilon(h\alpha)\varepsilon(g\alpha h) = -1$ because
both factors are fermionic.)

The examples suggest that $A^*_m = A_m$, $D^*_m = D_m$, and $B^*_m
= -C_m$.  To see that this is true for all $n$ one can write the
recursion formula in expanded form as
$$
\left(\begin{array}{cc} A_m  &-C_m \\ B_m  &D_m \end{array}\right)
= \left(\begin{array}{cc} g  &0 \\ 0 &-h\end{array}\right)
           \left(\begin{array}{cc} \beta  &-\alpha \\ -\alpha  &0
           \end{array}\right)
         \left(\begin{array}{cc} g  &0 \\ 0  &-h \end{array}\right) \cdots
     \left(\begin{array}{cc} \beta  &-\alpha \\ -\alpha  &0
     \end{array}\right)
      \left(\begin{array}{cc}  g  &0 \\  0  &-h \end{array}\right),
$$
which gives
$$
\left(\begin{array}{cc}  A_{m+2}  &-C_{m+2} \\ B_{m+2}  &D_{m+2}
\end{array}\right) = \left(\begin{array}{cc} g\beta  &-g\alpha \\
h\alpha  &0
\end{array}\right)
       \left(\begin{array}{cc}  A_m  &-C_m \\ B_m  &D_m
       \end{array}\right)
    \left( \begin{array}{cc} \beta g  &\alpha h \\ -\alpha g  &0
     \end{array}\right)
$$
and hence
$$
\begin{array}{l} A_{m+2} = g\beta A_m\beta g - g\alpha B_m\beta g + g\beta
C_m\alpha g
           + g\alpha D_m\alpha g \\
B_{m+2} = h\alpha A_m\beta g + h\alpha C_m\alpha g \\
C_{m+2} = g\alpha B_m\alpha h - g\beta A_m\alpha h \\
D_{m+2} = h\alpha A_m\alpha h.
\end{array}
$$
The desired conclusion $A^*_{m+2} = A_{m+2}$, $D^*_{m+2} =
D_{m+2}$, $B^*_{m+2} = -C_{m+2}$ now follows from the same
statement for $m$.

Combining the above we obtain
$$
\begin{array}{l}
A_m(t,x,y) = A_m(t,y,x) \\
B_m(t,x,y) = C_m(t,y,x) \\
D_m(t,x,y) = -D_m(t,y,x).
\end{array}
$$
With $G_m(t,x,y,\theta,\theta') = A_m + \theta B_m + \theta'C_m +
\theta\theta'D_m$ these three equations take the form
$$
G_m(t,x,y,\theta,\theta') = G_m(t,y,x,\theta',\theta).
$$
The same condition for the SUSY Green's function $G =
\ds\sum^\infty_0 G_m$ is the expression of its symmetry with
respect to $x \leftrightarrow y$ and $\theta \leftrightarrow
\theta'$ interchanges.

\section{The Asymptotic SUSY Green's Function}

\medskip
Computations of asymptotic series will be done in the following
setting.

Let $\Omega$ be an open subset of $\Bbb R^n$.  Then the space
$C^\infty_\uparrow (\ov\Omega)$ is to consist of the infinitely
differentiable functions defined in $\Omega$ which, together with
all partial derivatives, admit continuous extensions to the
closure of $\Omega$ and are bounded by polynomials.

When $\Omega \subset \Bbb R^3$ is the half--space $t > 0$, $L \in
C^\infty_\uparrow (\ov\Omega)$ admits the asymptotic expansion
$$
L(t,x,y) \sim \sum^\infty_{n=0} t^nL_n(x,y)
$$
in which $L_n(x,y) = \frac{1}{n!    } \ds\lim_{t\downarrow 0}
\partial^n_t L(t,x,y)$.  The error estimate is
$$
L(t,x,y) = \sum^N_{n=0} t^nL_n(x,y) + t^{N+1} R_{N+1}(t,x,y)
$$
where, by Taylor's Theorem $R_{N+1} \in C^\infty_\uparrow
(\ov\Omega)$.

Given $K \in C^\infty_\uparrow (\Bbb R^3)$, the square root
substitution may produce an element of
$C^\infty_\uparrow(\ov\Omega)$: if $K$ is defined everywhere in
$\Bbb R^3$ and satisfies $K(t,x,y) = K(-t,x,y)$ then $K(\sqrt
t,x,y) \in C^\infty_\uparrow(\ov\Omega)$.

When elements of these spaces appear in improper integrals the
convergence will be deduced from the polynomial growth conditions.
These permit separation of variables, in that every polynomial
satisfies
$$
|P(x_1,\ldots,x_n)| \le C(1+x_1^{2k})\cdots (1+x_n^{2k})
$$
for some $C > 0$, $k > 0$.

These preparations done, we examine the transformation $L
\rightarrow M$ given by
$$
g(t,x-y)M(t,x,y) = \int^t_0 \int^\infty_{-\infty} g(t-s,x-u)
        L(s,u,y)g(s,u-y)duds.
$$
The Gauss bell heat kernel admits the factorization
$$
g(t-s,x-u)g(s,u-y) = g(t,x-y)g(\tau,u-w)
$$
in which $0 < s < t$, $s = \sigma t$, $\tau = \sigma(1-\sigma)t$,
and $w = \sigma x + (1-\sigma)y$.  Thus, except for a constant
factor,
$$
\begin{array}{c} M(t,x,y) = \int^t_0 \int^\infty_{-\infty} L(s,u,y)
             g(\tau,u-w)duds \\
   = \int^t_0 \int^\infty_{-\infty} L(s,w+\sqrt\tau v,y)
          e^{-\frac{v^2}{4}} dvds
\end{array}
$$
after the change of variable $u = w+\sqrt\tau v$, $du = \sqrt\tau
dv$.

By hypothesis $L(t,x,y) \in C^\infty_\uparrow(\ov\Omega)$.  Then
$L(s,w+\tau v,y) \in C^\infty_\uparrow (\ov\Delta_1)$, where
$\Delta \subset \Bbb R^5$ is defined by $s > 0$.  The polynomial
bound gives the absolute convergence of
$$
J(s,\tau,w,y) = \int^\infty_{-\infty} L(s,w+\tau v,y)
          e^{-\frac{v^2}{4}} dv,
$$
a function in $C^\infty_\uparrow(\ov\Delta_2)$, with $\Delta_2
\subset \Bbb R^4$ defined by $s > 0$.

Evidently $J$ is invariant under $\tau \rightarrow -\tau$.
Therefore $J(s,\sqrt\tau,w,y) = K(s,\tau,w,y)$ is in
$C^\infty_\uparrow(\Delta_3)$, with $\Delta_3 \subset \Bbb R^4$
given by $s > 0$, $\tau > 0$.  This leaves $M$ in the form
$$
\begin{array}{c}
M(t,x,y) = \int^t_0 K(s,\tau,w,y)ds \\
   = t\int^1_0 K(\sigma t,\sigma(1-\sigma)t,\sigma x+(1-\sigma)y,y)
       d\sigma,
\end{array}
$$
after the change of variable $s = \sigma t$, $ds = td\sigma$.

It is now evident that the construction $L \rightarrow M$ is a
linear transformation into itself of the vector space
$C^\infty_\uparrow(\ov\Omega)$, with $\Omega \subset \Bbb R^3$
given by $t
>  0$ as before.  Furthermore if $L = \ds\sum_{n\ge N} t^nL_n(x,y)$ then
$M = \ds\sum_{n\ge N+1} t^nM_n(x,y)$.

When $\partial_x$ or $\partial_y$ is applied to the equation at
the very beginning, the integrand $gLg$ is replaced by $hLg$ or
$gLh$, and $M$ is replaced by $\partial_xM \pm \frac{x-y}{2t}M$,
for example.  This function is in $C^\infty_\uparrow(\Omega)$ once
more, but the cancellation of $t$ means that the resulting
asymptotic series may, like that of $L$, involve all $t^n$ with $n
\ge N$.

The Green's function $G = \ds\sum^\infty_{m=0} G_m$ can now be
expanded term by term.  In general, words $W =
e_0\omega_1e_1\omega_2 \cdots \omega_me_m$ give functions of three
variables $W(t,x,y)$ by the multiple integral formula. If $W$ is
lengthened to $e\omega W$ with $e = g$ or $h$ and $\omega =
\alpha$ or $\beta$, one has the recursion formula
$$
(e\omega W)(t,x,y) = \int^t_0 \int^\infty_{-\infty} e(t-s,x-u)
         \omega(u)W(s,u,y)duds.
$$
If $W(t,x,y) = g(t,x-y)L(t,x,y)$ for some $L \in C^\infty_\uparrow
(\ov\Omega)$ then $(e\omega W)(t,x,y) = g(t,x-y)M(t,x,y)$ where $M
\in C^\infty_\uparrow(\Omega)$ is the result of applying the $L
\rightarrow M$ transformation to $\omega(x)L(t,x,y)$ and then
taking a derivative if necessary.

The observations about lowest powers of $t$ permit the specific
formula
$$
W(t,x,y) = g(t,x-y) \sum^\infty_{n=N-1} t^nL_n(x,y)
$$
where $N = N(W)$ is the number of times that $g$ occurs in the
word $W$. Therefore, by induction, every word gives rise to an
asymptotic expansion as written, with coefficients $L_n(x,y) \in
C^\infty_\uparrow (\Bbb R^2)$.

Then, looking at $G_1$, we see that
$$
G_1(t,x,y,\theta,\theta') = g(t,x-y)\sum^\infty_{n=0} t^n
         \wh L_n(x,y,\theta,\theta').
$$
Each $\wh L_n$ is a linear combination of four $\Lambda$--valued
elements of $C^\infty_\uparrow (\Bbb R^2)$, with coefficients $1$,
$\theta$, $\theta'$, and $\theta\theta'$.  In particular $\wh
L_0(x,y,0,0)= 0$, because, with $A_1 = g\beta g$, we have
$A_1(0,x,y) = 0$.

When we substitute
$$
g(t,x-y) = G_0(t,x-y,\theta,\theta')
     \Big(1 - \frac{\theta\theta'(x-y)}{2t}\Big),
$$
no negative powers of $t$ appear, because $\theta\theta' \wh
L_0(x,y,\theta,\theta') = \theta\theta' A_1(0,x,y) = 0$.
Therefore
$$
G_1(t,x,y,\theta,\theta') =
G_0(t,x-y,\theta,\theta')\sum^\infty_{n=0}
      t^nL_n(x,y,\theta,\theta'),
$$
with coefficients in $C^\infty_\uparrow (\Bbb R^2)$ as before.

In the general case of $G_m$, the four entries in the matrix
recursion formula are finite linear combinations of words with
integer coefficients.  For such an entry, define $N$ to be the
minimum of $N(W)$ for the words appearing in the entry.

Suppose for some $m$ that $N(A_m) \ge p+1$ and $N(B_m)$, $N(C_m)$,
$N(D_m) \ge p$ for some nonnegative integer $p$.  This holds for
$m = 1$ and $m = 2$ with $p = 1$, by inspection of the formulas.
An examination of the $m \rightarrow m+2$ recursion formula then
shows that $N(A_{m+2}) \ge p+2$ while $N(B_{m+2}), N(C_{m+2}),
N(D_{m+2}) \ge p+1$.

This shows that
$$
\lim_{m\rightarrow\infty} N(G_m) = +\infty,
$$
with $N(G_m)$ being the minimum of $N(A_m),\ldots,N(D_m)$.  Thus,
with a change of notation,
$$
G_m(t,x,y,\theta,\theta') = G_0(t,x-y,\theta,\theta')\sum t^n
L_{m,n}(x,y,\theta,\theta').
$$
Passing to $\ds\sum^\infty_{m=0} G_m$ we note that only finitely
many $m$ can make a contribution to any given power of $t$.
Therefore
$$
G(t,x,y,\theta,\theta') \sim G_0(t,x-y,\theta,\theta')
\sum^\infty_{n=0}
       t^nM_n(x,y,\theta,\theta'),
$$
each $M_n$ being a combination of four elements of
$C^\infty_\uparrow (\Bbb R^2)$.  For example
$$
M_0 = 1 + \frac{\theta-\theta'}{2} \int^y_x \alpha(s)ds
      + \frac{\theta\theta'}{2} \int^y_x \beta (s)ds.
$$

The coefficients $M_n$ can be determined from a recursive system
of differential equations derived from the $G = G_0 \sum t^nM_n$
equation and the differential equation
$$
HG = D(\phi G).
$$
This completes the verification that the SUSY Green's function
possesses an asymptotic expansion, in the above sense.

In \cite{Adrian} it was shown that this asymptotic expansion,
which is now rigorously established, generated all the members of
the $N=1$ supersymmetric KdV hierarchy.

\section{Conclusions} A rigorous treatment has been given of the
Green's function $G$ of the $N=1$ supersymmetric heat equation in
one spatial dimension with a distribution initial value. The
convergence of the series defining $G$ is proven as well as the
existence of its asymptotic expansion as $t\rightarrow0^+$. It was
shown in \cite{Adrian} that the formal asymptotic expansion
generates all the members of the supersymmetric $N=1$ KdV
hierarchy. The asymptotic expansion for the Green's function of
the KdV equation was discussed in \cite{Avramidi}.

\end{document}